\documentclass[11pt]{article}
\def\va{Virasoro algebra}

\newcommand{\LL}{Lanczos-Lovelock }
\def\eq#1{{Eq.~(\ref{#1})}}

\usepackage{amssymb}
\usepackage{epsfig}
\usepackage[active]{srcltx}
\begin{document}
\title{{\bf{\Large Noether Current, Horizon Virasoro Algebra and  Entropy}}}
\author{ 
 {\bf {\normalsize Bibhas Ranjan Majhi}$
$\thanks{E-mail: bibhas@iucaa.ernet.in}} \, and
{\bf {\normalsize T. Padmanabhan}$
$\thanks{E-mail: paddy@iucaa.ernet.in}}\\
 {\normalsize IUCAA, Post Bag 4, Ganeshkhind,}
\\{\normalsize Pune University Campus, Pune 411 007, India}
\\[0.3cm]
}

\maketitle

\begin{abstract}
We provide a simple and straightforward procedure for defining a Virasoro algebra based on the diffeomorphisms near a null surface in a spacetime and obtain the entropy density of the null surface from its central charge. We use the \textit{off-shell} Noether current corresponding to the diffeomorphism invariance of a  gravitational Lagrangian $L(g_{ab},R_{abcd})$  and define the \va\ from its variation. This allows us to identify  the central charge and the zero mode eigenvalue using which we obtain the entropy density of the  Killing horizon. Our approach works for all \LL\ models and reproduces the correct Wald entropy. The entire analysis is done off-shell without using the field equations and allows us to define an entropy density for any null surface which acts as a local Rindler horizon for a particular class of observers.
\end{abstract}

\section{Introduction and Motivation}

Several recent results strongly indicate the possibility that the field equations of gravity have the same status as the equations of fluid mechanics or elasticity. (For a recent review, see Ref. \cite{paddyaspects}.)
This approach has a long history originating from the work of Sakharov \cite{sakharov} and interpreted in many ways by different authors (for a incomplete sample of references, see Ref. \cite{others}). One specific implementation of this idea considers the \textit{field equations} of the theory to be  `emergent' in a well-defined sense, rather than use that term in a more speculative vein --- like e.g., considering the space and time themselves to be emergent etc.
The evidence for such a specific interpretation comes from different facts like the possibility of interpreting the field equation in a wide class of theories as thermodynamic relations \cite{thermo}, the nature of action functional in gravitational theories and their thermodynamic interpretation \cite{holoaction}, the possibility of obtaining the field equations from a thermodynamic extremum principle \cite{aseemtp}, application of equipartition ideas to obtain the density of microscopic degrees of freedom \cite{equiv}, the equivalence of Einstein's field equations to the Navier-Stokes equations near a null surface \cite{NS} etc. 
Two key features of this paradigm are the following. 
\begin{itemize}
\item 
First, these results show that the connection between gravitational dynamics and horizon thermodynamics is quite deep and goes well beyond Einstein's theory of gravity. It seems to have its roots in general covariance and principle of equivalence which  allows one to introduce the concept of  local Rindler observers and local Rindler horizons in the neighbourhood of any event in the spacetime. (For a conceptual description of this point of view, see Ref. \cite{dialog}.) In fact, the strongest theoretical evidence for such an emergent paradigm is the fact that  the results related to the thermodynamics of the horizons in the context of GR generalizes in a natural fashion, to a much wider class of theories like \LL\ models. 
\item
Second, it appears to be possible and useful to attribute an observer dependent entropy density to any null surface rather than to horizons which arise as solutions to the field equations. The entropy density (in contrast to the temperature) knows the underlying field equations and, in fact, the field equations can be obtained by extremizing a suitably defined entropy density of spacetime.
\end{itemize}

The emergent paradigm, therefore, motivates us to study all the conventional  approaches to the derivation of horizon entropy from a broader perspective.
 We know from the pioneering works of Bekenstein \cite{Bekenstein:1973ur} and Hawking \cite{Hawking:1974rv} that, in GR, one can attribute an entropy $S=(1/4)A$ to a black hole horizon where $A$ is the horizon area. In the decades following the original derivation, several alternative approaches have led to the same expression for black hole entropy in GR; (see, for a nonrepresentative sample, Refs. \cite{Strominger:1996sh,Ashtekar:1997yu,Bombelli:1986rw}). But we should keep in mind  the fact that area has a simple geometrical meaning which allows one to obtain the horizon entropy in GR (once we know the result!) in several different ways creating an impression of ``universality''
for this result. 

On the other hand, the proportionality between entropy and area does not hold for more general class of gravitational theories in which the entropy is given by a prescription due to  Wald \cite{Wald:1993nt} which essentially identifies the horizon entropy with a suitably defined Noether charge. 
Many of the approaches which correctly reproduces $S\propto A$ in the context of GR cannot be generalized in a natural fashion to more general class of theories like, for example, \LL\ models. (One such  example, which does \textit{not} generalise, is the entanglement entropy; see, for example, Ref. \cite{TPentangle}.)
\textit{Therefore, the possibility of generalization beyond GR acts as an acid test in discriminating between the different approaches for obtaining the horizon entropy.} 
Further, in the context of emergent paradigm, the entropy density of a null surface is a primitive concept which should not require one to use the field equations.  In other words, one should be able to derive the expression for entropy using only off-shell constructs and in a manner applicable to any local Rindler horizon. 
In summary, we expect valid procedures for obtaining the entropy of the horizon to satisfy the following three criteria: 
\begin{itemize}
\item 
It should be based on off-shell constructs which are independent of the field equations.
\item
One should be able to use the procedure to obtain the entropy density of any null surface which can act as a local Rindler horizon        rather than be specific to certain solutions to the field equations like black holes.
\item
The method should work for all \LL\ models of gravity rather than be specific to Einstein's theory.
\end{itemize}
Past work has shown that many of the usual derivations does satisfy these criteria which formed the basis for the claim that the gravity-thermodynamics connection transcends GR. 

In this paper, we explore yet another approach to horizon entropy explored in the literature, mainly in the context of black hole horizons, using the existence of a \va\ and central charge in these spacetimes.
This is based on an approach  by Brown and Henneaux initiated by \cite{Brown:1986nw}, which was originally used in the context of the ($1+2$) dimensional gravity with asymptotically AdS space-time. They found  that the Fourier modes $Q_j$ of the charges corresponding to the asymptotic diffeomorphism symmetry generators obey Virasoro algebra with central extension:
\begin{eqnarray}
i[Q_m,Q_n] = (m-n)Q_{m+n}+\frac{C}{12}m^3\delta_{m+n,0}~,
\label{virasoro}
\end{eqnarray}
where $C$ is known as the central charge. The work by Strominger \cite{Strominger:1997eq} and others showed that
 if one uses the above central charge in the Cardy formula \cite{Cardy:1986ie,Carlip:1998qw},
the resulting entropy comes out to be the Bekenstein-Hawking entropy  for the ($1+2$) dimensional black hole. 
This was further developed by Carlip  \cite{Carlip:1998wz,Carlip:1999cy} using the diffeomophism symmetry generators near the horizon
to lead to the black hole entropy. In this approach, one begins with the  diffeomorphism generators $\xi_n^a(x)$ which preserve a set of boundary conditions near the horizon. The Fourier modes of these generators obey one sub-algebra isomorphic to Diff $S^1$ given by
\begin{eqnarray}
i\{\xi_m,\xi_n\}^a = (m-n)\xi^a_{m+n}
\label{Virasoro1}
\end{eqnarray}
where  $\{,\}$ is the Lie bracket. One can then construct the Fourier  modes $Q_n$  of the charges  corresponding to each $\xi^a_n$, either by Hamiltonian \cite{Regge:1974zd} or a covariant Lagrangian formalism \cite{Katz:1996nr,Wald:1993nt,Julia:1998ys,Julia:2000er,Silva:2000ys} and evaluate the Lie brackets among them. A comparison between this algebra and (\eq{virasoro}) allows us to identify the central charge. Finally, one finds the zero mode eigenvalue $Q_0$ and computes the entropy of the black hole using Cardy formula. 
Several related approaches have been developed using these ideas
with the hope that diffeomorphism symmetry generators may shed some light towards the microscopic degrees of freedom responsible for entropy of the horizon \cite{Solodukhin:1998tc,Brustein:2000fw,Lin:1999gf,Das:2000zs,Jing:2000yn,Terashima:2001gn,Koga:2001vq,Park:2001zn,Carlip:2002be,Cvitan:2002cs,Cvitan:2002rh,Park:1999tj,Park:1999hs,Dreyer:2001py,Silva:2002jq}.
All these approaches developed in the literature have the following ingredients: 
\begin{itemize}
\item 
The Noether current used in the approaches are defined on-shell, usually by ignoring a term which vanishes when equations of motion are used.
\item
The calculation of Lie brackets is somewhat complicated and different approaches lead to slightly different results and it is often not clear how to interpret these differences in the calculations. 
\item
To obtain the correct result, one often has to impose specific boundary conditions on the horizon in order to set certain terms to zero. Again, the physical meaning of these boundary conditions is often not clear.
\item
Most of the analysis (except the one in \cite{Cvitan:2002rh}) is confined to GR and it is not clear how to generalize the results for a wider class of theories.
\end{itemize}
In this paper, we revisit this approach to horizon entropy and show that there is a relatively simple way of obtaining the central charge and the horizon entropy using the off-shell Noether current in any generally covariant theory of gravity.  The procedure, for example, works in a straightforward manner for any \LL\ model and does not require us to impose any  boundary conditions to make unwanted terms to vanish. The essential idea is to use the diffeomorphism invariance  of the Lagrangian under  $x^i \to x^i + \xi^i_1$ to define a Noether current $J_a[\xi_1]$  and then use \textit{its} variation $\delta_{\xi_2} J_a[\xi_1]$ under a second diffeomorphism $x^i \to x^i + \xi^i_2$ to define the Lie bracket structure.  This can be done without using the explicit form of the Noether current and thus works for a wide class of gravitational theories. The calculation of the resulting Lie bracket is quite simple algebraically and leads to the standard results without us having to impose any extra boundary conditions.  One can then identify the resulting \va, the central charge and the zero-mode eigenvalue. Using these in the Cardy formula leads to the entropy density of the null surface which turns out to be the same as Wald entropy.

We summarise below the key new features of this paper:
\begin{itemize}
\item The current considered here is \textit{defined and conserved} off-shell i.e. we do not use the  equations of motion in its definition and it is conserved identically {\footnote{A complete discussion on the derivation of noether current and its conservation for \textit{on-shell} condition is given in \cite{New1}. In this case a suitable boundary condition is required to obtain the necessary results. But the present paper, as we shall see, will deal with off-shell situation.}}. This is in contrast to --- and an improvement on --- the previous approaches which use equations of motion extensively.  

\item The definition of the bracket among the charges is completely general in the sense that  one does not need the explicit expression for the noether current to obtain its structure.

\item  We provide a discussion of the derivation of the central term (and zero mode energy) from the surface term contribution $\delta v^a$ in the Noether current. This is completely new and does not seem to have been noticed in the previous literature.

\item  In the derivation of bracket we do not use the  equations of motion or specific  boundary conditions (like Dirichlet or the Neumann boundary condition) to make noncovariant terms vanish; therefore  our approach is  general enough to include any covariant theory of gravity. (This feature was missing in the earlier literature which used the equations of motion for the evaluation of the relevant brackets as an on-shell construct.) As a direct consequence of the above, the central term we obtain is also an off-shell construct.

\end{itemize}

The last comment about boundary conditions requires some clarifications. To begin with, for given Noether current $J^a$, one can assign several Noether potentials $J^{ab}$ differing by addition of  divergence-free antisymmetric tensors.  In our approach, we can take $J^{ab}\propto P^{abcd}\nabla_c\xi_d$ as the fundamental quantity given to us and define the charge as an integral over the stretched horizon of $J^{ab}d\Sigma_{ab}$. This is well-defined, when we start from a specific Noether potential $J^{ab}$ (rather than Noether current). We then only need to assume that the corresponding integral in the outer boundary (say, at asymptotically large distance) vanishes to relate integrals over $J^{ab}d\Sigma_{ab}$ to integrals of $J^a$ in the bulk. So we do need an asymptotic boundary condition which is almost always assumed in such discussions. 
\textit{But in the literature one often finds the use of much less motivated and possibly more stringent additional boundary conditions to throw away terms which arise in the computation.} 
For example, a
derivation of the bracket among the noether charges for the $\Gamma^2$ term in Einstein-Hilbert action has been discussed in \cite{Silva:2002jq}. Here the noether potential comes out to be the Katz potential (see the derivation of \cite{Julia:1998ys} in section 5.1), which is not in covariant form. Hence the variation of the potential will contain some terms which are not again covariant. To deal this issue,  one needs to impose some boundary condition such that the non-covariant terms do not appear in the final expression. (For instance, see the general discussion around Eq. (4) of the ref. \cite{Silva:2000ys}.) Here, the variation of the potential is given by the general expression in Eq. (4) which is `integrable' only under suitable boundary condition. \textit{In particular, for $\Gamma^2$ Lagrangian, the Dirichlet or the Neumann boundary condition has been used} \cite{Julia:2000er}. Similar situation arises in some other approaches like the background metric method \cite{Deruelle:2003ps,Katz:1996nr}.  A comment  in \cite{Julia:2000er} (see at the beginning paragraph of section 2.2) says that the use of the background metric is nothing but a covariant way to impose the \textit{required boundary conditions}. Thus all these approaches require extra boundary conditions to get the correct result.
In our approach, on the other hand, since our noether current/potential is manifestly covariant, we do not need any of the above mentioned boundary conditions. In this sense our method does not care about these boundary conditions, described above.

The plan of the paper is as follows: 
We begin in section 2 by computing the variation of Noether current under diffeomorphism and use it in section 3 to define a suitable bracket of the charges. We also point out the differences between our approach and the previous ones in the literature in section 3 and in section 4. Section 5 uses Carlip's procedure to evaluate the horizon entropy in a general \LL\ model of gravity
using the Cardy formula. The details of the computation are given in Appendix \ref{AppendixB} since we could not find some of these explicitly done in published literature. The entire procedure is illustrated using the local Rindler horizon around an event in an arbitrary spacetime in section 6 and section 7 gives the conclusions.

\section{Variation of Noether charge under diffeomorphism}

For a generally covariant Lagrangian, the conserved Noether current $J^a$ can be expressed as the covariant derivative of an anti-symmetric tensor $J^{ab}$ called the super-potential with a corresponding current density $P^a$. These quantities satisfy the standard conservation laws which are valid off-shell:
\begin{eqnarray}
J^a \equiv \nabla_bJ^{ab}, \quad P^a \equiv \sqrt{g}J^a; \quad  \quad \nabla_a J^a =0, \quad \partial_a P^a =0~.
\label{1.07}
\end{eqnarray}
 Let us now consider the variation of the current density itself for an arbitrary diffeomorphism $x^a\rightarrow x^a+\xi^a$. We have:
\begin{eqnarray}
\delta_{\xi} P^a &\equiv& \pounds_{\xi}P^a = \sqrt{g}\Big[ J^a\nabla_b\xi^b + \xi^b\nabla_bJ^a - J^b\nabla_b\xi^a\Big]
\nonumber
\\
&=& \sqrt{g} \Big[\nabla_b(\xi^b J^a - \xi^a J^b) + \xi^a \nabla_bJ^b\Big] = \sqrt{g} \Big[\nabla_b(\xi^b J^a - \xi^a J^b)\Big]
\label{1.08}
\end{eqnarray}
because $\nabla_b J^b =0$.
 The variation of the corresponding Noether charge is defined as:
\begin{eqnarray}
\delta_\xi Q \equiv \int d\Sigma_a \delta_{\xi} P^a = \int d\Sigma_{ab}\sqrt{h}\xi^b J^a~,
\label{1.119}
\end{eqnarray}
where $h$ is the determinant of the induced metric of ($d-2$)-dimensional boundary in a $d$-dimensional spacetime.

    The same result  can also be obtained from the variation of the super-potential density $P^{ab} \equiv \sqrt{h}J^{ab}$ in the following way:
\begin{eqnarray}
\delta_{\xi} P^{ab} &\equiv&  \pounds_\xi P^{ab}=  \sqrt{h}\Big[J^{ab}\nabla_c\xi^c + \xi^c\nabla_cJ^{ab} - J^{cb}\nabla_c\xi^a - J^{ac}\nabla_c\xi^b\Big]
\nonumber
\\
&=& \sqrt{h}\Big[\nabla_c(\xi^cJ^{ab}) - J^{cb}\nabla_c\xi^a - J^{ac}\nabla_c\xi^b\Big]
\nonumber
\\
&=& \sqrt{h}\Big[\xi^bJ^a - \xi^aJ^b \Big] + \sqrt{h}\nabla_c\Big[\xi^aJ^{bc} + \xi^bJ^{ca} - \xi^cJ^{ba} \Big]~.
\label{1.11}
\end{eqnarray}
The first term  leads to \eq{1.119} on integration while the second term, after integrating over the ($d-2$)-surface, can be expressed as,
\begin{eqnarray}
\int d\Sigma_{ab} \sqrt{h}\nabla_c\Big[\xi^aJ^{bc} + \xi^bJ^{ca} + \xi^cJ^{ab} \Big] = 2\int d\Sigma_{a} \sqrt{g}\nabla_b\nabla_c\Big[\xi^aJ^{bc} + \xi^bJ^{ca} - \xi^cJ^{ba} \Big] =0
\label{1.120}
\end{eqnarray}
where the last equality follows from the fact that the expression in square brackets is antisymmetric in $b,c$ and is equal in the cyclic permutation of all the three indices.
This shows that, although the variation of the super-potential possesses an extra term, both the current and the super-potential lead to same expression for  the variation of the charge.

\section{Virasoro algebra and the central term}

We will now define a (Lie)   bracket structure for the charges and show that it leads to the  usual Virasoro algebra with the central extension. The analysis will be done for a general Lagrangian of the kind $L=L(g_{ab},R_{abcd})$.
Let us define the relevant bracket among the charges as: 
\begin{equation}
[Q_1,Q_2] \equiv (\delta_{\xi_1} Q[\xi_2] - \delta_{\xi_2} Q[\xi_1])                                                                 \end{equation} 
Then using Eq.~(\ref{1.119}), we obtain the general expression,
\begin{eqnarray}
[Q_1,Q_2]: = \int \sqrt{h}d\Sigma_{ab}\Big[\xi^a_2J^b[\xi_1] - \xi^a_1J^b[\xi_2] \Big]
\equiv\int \sqrt{h}d\Sigma_{ab}\Big[\xi^a_2J^b_1 - \xi^a_1J^b_2 \Big]
\label{1.78}
\end{eqnarray}
where we use the notation $J^b_1=J^b[\xi_1]$ etc.
We note that: 
(a) This definition is quite general and has not used any field equations. 
(b) The derivation of \eq{1.78} is simple and straightforward. 
(c) If we use the form $J^a = \nabla_b(\nabla^a\xi^b - \nabla^b\xi^a)$ for GR one can easily obtain the result obtained earlier in the literature, like e.g., in  Ref. \cite{Carlip:1999cy}.
We will hereafter concentrate  on Lanczos-Lovelock gravity and evaluate  Eq. (\ref{1.78}) on the ($d-2$)-dimensional null surface which is a Killing horizon.

 A definition of the Lie bracket is also given in \cite{Silva:2002jq} using similar ideas. The main differences between our definition (\ref{1.78}) and that given in \cite{Silva:2002jq} are the following:  (i)  In \cite{Silva:2002jq} the Lagrangian considered was the $\Gamma\Gamma-\Gamma\Gamma$ part of the Einstein-Hilbert action \cite{Julia:1998ys,Julia:2000er,Silva:2002jq} which is not covariant and hence the Noether current is not an tensor. In our case we start with a scalar action and the resulting expressions are tensorial. (ii) In the previous work, while calculating the variation,  of the charge the variation of the diffeomorphism parameter $\xi$ was set to zero, i.e. $\delta_{\xi_1}\xi_2=0$ \cite{Silva:2002jq} and only the variation of the metric was retained. This is ill defined because  $\delta_{\xi_1}\xi_2 \equiv \pounds_{\xi_1}\xi_2^a =  \{\xi_1,\xi_2\}^a = 0$ which contradicts (\ref{Virasoro1}). In our approach, we use the usual definition of the Lie derivative and hence this difficulty is automatically avoided.

To evaluate  Eq. (\ref{1.78}) over the Killing horizon, we will follow the `stretched horizon' approach of Carlip \cite{Carlip:1999cy}. Let us first mention some of the key results needed for this computation. The location of the horizon is defined by the vanishing of the norm of a timelike (approximate) Killing vector $\chi^a$. Near the horizon, one can define a vector $\rho^a$, orthogonal to the orbits of the Killing vector $\chi^a$, by the following relation
\begin{eqnarray}
\nabla_a\chi^2=-2\kappa\rho_a~,
\label{1.121}
\end{eqnarray}
where $\kappa$ is the surface gravity at the horizon, with $\chi^a\rho_a=0$. Consider a class of diffeomorphism generators given by:
\begin{eqnarray}
\xi^a = T\chi^a + R\rho^a~.
\label{1.122}
\end{eqnarray}
where $T$ and $R$ are scalar functions chosen such that the generators obey the (near-horizon) condition
$[(\chi^a\chi^b)/\chi^2]\delta_{\xi} g_{ab}\rightarrow 0$ which preserves the 
horizon structure.  This condition leads to a relation among $R$ and $T$ given by:
\begin{eqnarray}
R= \left( \frac{\chi^2}{\rho^2 \kappa}\right) \, \chi^a \nabla_a T \equiv\frac{\chi^2}{\kappa\rho^2}DT~,
\label{1.123}
\end{eqnarray}
where $D\equiv \chi^a\nabla_a$. The diffeomorphism characterised by Eq.~(\ref{1.122}) and Eq.~(\ref{1.123}) form a closed sub-algebra if
\begin{eqnarray}
\rho^a\nabla_aT = 0~,
\label{1.124}
\end{eqnarray}
near the horizon. Later on in Section \ref{Section6}, we will demonstrate this explicitly by using the Rindler metric in the Riemann normal coordinates and show that the above condition is exact upto ${\cal{O}}(\chi^2)$.

    For the Lanczos-Lovelock gravity the expression for $J^a$  and $J^{ab} $ are given by (see, e.g.,\cite{Padmanabhan:2010zzb}),
 \begin{eqnarray}
J^a = \frac{1}{8\pi G}P^{abcd}\nabla_b\nabla_c\xi_d~, \quad J^{ab} = \frac{1}{8\pi G}P^{abcd}\nabla_c\xi_d
\label{1.79}
\end{eqnarray}
where $P^{abcd}=(\partial L/\partial R_{abcd})$ is called the entropy tensor of the \LL theory. This tensor has the same algebraic symmetries of the curvature tensor, viz. it is anti-symmetric in $a,b$ and $c,d$; symmetric in the interchange of the pairs ($a,b$) and ($c,d$) and $P^{a(bcd)} = 0$.
Near the Killing horizon, this expression for the Noether current reduces to the following form:
\begin{eqnarray}
J^a = \frac{1}{8\pi G}P^{abcd}\chi_b\chi_c\rho_d \frac{1}{\chi^4} (2\kappa DT - \frac{1}{\kappa}D^3T) + {\cal{O}}(\chi^2)~.
\label{1.80}
\end{eqnarray}
when we use \eq{1.76} of Appendix \ref{AppendixB}. Since the later analysis will be done near the null surface where $\chi^2\rightarrow 0$, the terms from ${\cal{O}}(\chi^2)$ will be neglected and will not be mentioned explicitly. (An illustration of this fact will be provided in the case of Rindler geometry in Section \ref{Section6}.)
Now, the surface element $d\Sigma_{ab}=d^{d-2}X\mu_{ab}$ is given by Eq.~(\ref{1.81}) of Appendix  \ref{AppendixA}.
Therefore,
\begin{eqnarray}
d\Sigma_{ab} \xi_2^aJ_1^b = -(d^{d-2}X) \frac{1}{8\pi G}\frac{|\chi|}{\rho\chi^4}P^{becd}\rho_b\chi_c\rho_d\chi_e (2\kappa DT_1 - \frac{1}{\kappa}D^3T_1)T_2~.
\label{1.82}
\end{eqnarray}
But,
\begin{eqnarray}
P^{becd}\mu_{be}\mu_{cd} = \frac{4}{\rho^2\chi^2}P^{becd}\rho_b\chi_c\rho_d\chi_e~.
\label{1.83}
\end{eqnarray}
Substituting this in Eq.~(\ref{1.82}) we obtain,
\begin{eqnarray}
d\Sigma_{ab} \xi_2^aJ_1^b = d^{d-2}X\frac{1}{32\pi G}\frac{\rho}{|\chi|}P^{abcd}\mu_{ab}\mu_{cd}(2\kappa DT_1 - \frac{1}{\kappa}D^3T_1)T_2~.
\label{1.84}
\end{eqnarray}
Hence the bracket Eq.~(\ref{1.78}) evaluates near the horizon,  to,
\begin{eqnarray}
[Q_1,Q_2]: =&& \frac{1}{32\pi G}\int \sqrt{h}d^{d-2}XP^{abcd}\mu_{ab}\mu_{cd}
\nonumber
\\
&&\times\Big[\frac{1}{\kappa}(T_1 D^3T_2 - T_2 D^3T_1) - 2\kappa(T_1DT_2 - T_2DT_1)\Big]
\label{1.85}
\end{eqnarray}

We will next obtain an expression for the charge $Q[\xi]$ in the near horizon limit:
\begin{eqnarray}
Q[\xi] =  \frac{1}{2}\int d\Sigma_{ab}\sqrt{h}J^{ab}~.
\label{1.86}
\end{eqnarray}
For Lanczos-Lovelock gravity  the expression for $J^{ab}$ in  \eq{1.79} becomes, on using  Eq.~(\ref{1.55}) of  Appendix \ref{AppendixB} to  the following form:
\begin{eqnarray}
J^{ab} = \frac{1}{8\pi G}P^{abcd}\,\chi_c\rho_d\, \left[\frac{2\kappa}{\chi^2}T - \frac{1}{\kappa\chi^2}D^2T\right]~.
\label{1.87}
\end{eqnarray}
Next we will show for a Rindler metric in Riemann normal coordinates that the above relation is exact upto order ${\cal{O}}(\chi^2)$.
Since,
\begin{eqnarray}
P^{abcd}\mu_{cd} = -\frac{2|\chi|}{\rho\chi^2}P^{abcd}\chi_c\rho_d~,
\label{1.88}
\end{eqnarray}
we get,
\begin{eqnarray}
J^{ab} = -\frac{1}{16\pi G}\frac{\rho}{|\chi|}P^{abcd}\mu_{cd}\Big[2\kappa T - \frac{1}{\kappa}D^2T\Big]
\label{1.89}
\end{eqnarray}
leading to:
\begin{eqnarray}
Q[\xi] &=&  -\frac{1}{32\pi G}\int\sqrt{h}d^{d-2}X\mu_{ab}\frac{\rho}{|\chi|}P^{abcd}\mu_{cd}\Big[2\kappa T - \frac{1}{\kappa}D^2T\Big]
\nonumber
\\
&=& -\frac{1}{32\pi G}\int\sqrt{h}d^{d-2}XP^{abcd}\mu_{ab}\mu_{cd}\Big[2\kappa T - \frac{1}{\kappa}D^2T\Big]
\label{1.90}
\end{eqnarray}

 We can now compute the central term  defined by the relation,
\begin{eqnarray}
K[\xi_1,\xi_2] = [Q_1,Q_2] - Q[\{\xi_1,\xi_2\}]~,
\label{1.77}
\end{eqnarray}
where $[Q_1,Q_2]$ is given by Eq.~(\ref{1.85}) and $Q[\{\xi_1,\xi_2\}]$ will be obtained by using Eq.~(\ref{1.90}). 
The Lie bracket $\{\xi_1,\xi_2\}^a$, near the horizon is:
\begin{eqnarray}
\{\xi_1,\xi_2\}^a &=& \xi_1^b\nabla_b\xi^a_2  - \xi_2^b\nabla_b\xi_1^a
\nonumber
\\
&=& (T_1DT_2 - T_2DT_1)\chi^a - \frac{1}{\kappa}D(T_1DT_2-T_2DT_1)\rho^a
\nonumber
\\
&\equiv& \{T_1,T_2\}\chi^a + \{R_1,R_2\}\rho^a~,
\label{1.91}
\end{eqnarray}
where $\{T_1,T_2\} = (T_1DT_2 - T_2DT_1)$, etc.
Using this in Eq.~(\ref{1.90}) we get:
\begin{eqnarray}
Q[\{\xi_1,\xi_2\}] &=& - \frac{1}{32\pi G}\int\sqrt{h}d^{d-2}XP^{abcd}\mu_{ab}\mu_{cd}\Big[2\kappa \{T_1,T_2\} - \frac{1}{\kappa}D^2(\{T_1,T_2\})\Big]
\nonumber
\\
&=&- \frac{1}{32\pi G}\int\sqrt{h}d^{d-2}XP^{abcd}\mu_{ab}\mu_{cd}\Big[2\kappa (T_1DT_2 - T_2DT_1) 
\nonumber
\\
&&- \frac{1}{\kappa}(DT_1~D^2T_2 + T_1D^3T_2 - DT_2~D^2T_1 - T_2D^3T_1)\Big]
\label{1.92}
\end{eqnarray}
Therefore, substituting \eq{1.85} and \eq{1.92} in  Eq.~(\ref{1.77}), we obtain the central term to be:
\begin{eqnarray}
K[\xi_1,\xi_2] = - \frac{1}{32\pi G}\int\sqrt{h}d^{d-2}XP^{abcd}\mu_{ab}\mu_{cd}\frac{1}{\kappa}\Big[DT_1~D^2T_2 - DT_2~D^2T_1\Big]~.
\label{1.93}
\end{eqnarray}
This was obtained earlier in \cite{Cvitan:2002rh} by symplectic current (or potential) technique and using the on-shell expressions. Here, we derived this using  Eq.~(\ref{1.78}) without using the field equations or any special boundary conditions thereby demonstrating the generality of the result.

\section{Aside: Central term  from   the surface term contribution to  the Noether current}

We know that, the variation of a generally covariant Lagrangian of the from  $L(g_{ab},R_{abcd})$ under the variation $g^{ab}\to g^{ab}+\delta g^{ab}$ is given by the generic expression
\begin{equation}
\delta(L\sqrt{g})=\sqrt{g}[E_{ab}\delta g^{ab}+\nabla_a\delta v^a]
\end{equation} 
where the first term leads to the equation of motion of the form $E^{ab}=(1/2) T^{ab}$ while the second term is a surface contribution. When $\delta g^{ab}$ arises due to a diffeomorphism $x^i\to x^i+\xi^i$, this leads to an off-shell conservation law for the  corresponding conserved Noether current (see e.g., \cite{Padmanabhan:2010zzb}) given by:
\begin{equation}
J^a = \frac{1}{16\pi G} \Big( 2E^a_b\xi^b + L\xi^a \Big) + \delta_{\xi}v^a 
\equiv \frac{2}{16\pi G}\mathcal{R}^a_b\xi_b + \delta_{\xi}v^a  
=\frac{1}{8\pi G} P^{abcd}\nabla_b\nabla_c\xi_d 
\end{equation} 
where the last equality holds for the 
 \LL\ models. The consistency of these expressions arises from the fact that for \LL models: 
 \begin{equation}
\mathcal{R}^a_b=P^{aijk}R_{bijk}; \quad \delta v^a = -\frac{2}{16\pi G} P^{abcd}\nabla_d\Big(\nabla_b\xi_c+\nabla_c\xi_b\Big). 
\end{equation} 
So we can express the boundary term as:
\begin{equation}
\sqrt{g}\delta_{\xi}v^a=\sqrt{g}J^a - \sqrt{g}M^a;\quad
M^a = \frac{2}{16\pi G}{\mathcal{R}}^a_b\xi^b = \frac{2}{16\pi G} P^{aijk}R_{bijk}\xi^b
\end{equation} 

Before proceeding further, we will point out a curious result: We can obtain the same results
for the charge, central term etc, obtained in the previous section  by using the contribution from the boundary term $\delta v^a$ instead of $J^a$.
     To show this, let us start, as before, with the current density contributed only by the surface term 
\begin{eqnarray}
P^a_{\mathrm{surf}}[\xi]=\sqrt{g}\delta_\xi v^a=\sqrt{g}J^a - \sqrt{g}M^a
\label{1.98}
\end{eqnarray}
and compute its variation
\begin{eqnarray}
\delta_{\xi_1} (\sqrt{g}\delta_{\xi_2} v^a) = \sqrt{g}\nabla_b\Big[(\delta_{\xi_2}v^a) \xi_1^b - (\delta_{\xi_2}v^b) \xi_1^a\Big] + \sqrt{g} \xi^a_1 \nabla_b(\delta_{\xi_2}v^b)~,
\label{1.20}
\end{eqnarray}   
which yields,
\begin{eqnarray}
\delta_{\xi_1}P^a_{\mathrm{surf}} [\xi_2] &=& \delta_{\xi_1}(\sqrt{g}\delta_{\xi_2} v^a) = \sqrt{g} \nabla_b \Big[\xi_1^b (J^a_2-M^a_2) - \xi^a_1 (J^b_2-M^b_2)\Big] + \sqrt{g}\xi_1^a\nabla_b(J_2^b-M^b_2)
\nonumber
\\
&=& \sqrt{g}\nabla_b(\xi_1^bJ_2^a - \xi^a_1J_2^b) - \sqrt{g}\nabla_b(\xi_1^bM_2^a) + \sqrt{g}M_2^b\nabla_b\xi_1^a
\nonumber
\\
&=& \delta_{\xi_1}P^a[\xi_2] - \sqrt{g}\nabla_b(\xi_1^bM_2^a) + \sqrt{g}M_2^b\nabla_b\xi_1^a~.
\label{1.99}
\end{eqnarray}
Integrating over the ($d-1$) surface we obtain,
\begin{eqnarray}
\int d\Sigma_a \delta_{\xi_1}P^a_{\mathrm{surf}} [\xi_2] 
&=& \int d\Sigma_a \delta_{\xi_1}P^a[\xi_2] - \int d\Sigma_a \sqrt{g}\nabla_b(\xi_1^bM_2^a) + \int d\Sigma_a\sqrt{g}M_2^b\nabla_b\xi_1^a\nonumber\\
&=&\int d\Sigma_a \delta_{\xi_1}P^a[\xi_2] - \frac{1}{2}\int d\Sigma_{ab} \sqrt{h}\xi_1^bM_2^a + \int d\Sigma_{a}\sqrt{g}M_2^b\nabla_b\xi_1^a
\label{1.100}
\end{eqnarray}
Near the horizon (\ref{1.76}) yields.
\begin{eqnarray}
&&P^{abcd} R_{bdce}\xi^e = P^{abcd}(\nabla_b\nabla_d\xi_c - \nabla_d\nabla_b\xi_c) = 0~;
\nonumber
\\
&&P^{abcd} R_{bedc}\xi^e = 2P^{abcd}\nabla_d\nabla_c\xi_b = 0~.
\label{1.101}
\end{eqnarray}
Therefore, $M^a = 0$ and hence near the event horizon (\ref{1.98}) and (\ref{1.100}) lead to the required forms (\ref{1.86}) and (\ref{1.78}) respectively. Rest of the steps are identical to those in the previous section and leads to the same central term (\ref{1.93}).

The possible relevance of this result is as follows:
In obtaining the conserved Noether current, one usually uses the diffeomorphism invariance of the Lagrangian which allows one to write $\delta_\xi (L\sqrt{-g})$ as a four-divergence. The fact that $L$ is a scalar is \textit{sufficient} condition for $\delta_\xi (L\sqrt{-g})$ to be a total divergence but it is not a \textit{necessary} condition. There are Lagrangians (like the $\Gamma^2$ in GR) which are \textit{not} generally covariant scalars but still leads to an expression for $\delta_\xi (L\sqrt{-g})$ which is a  total divergence. Obviously, even such Lagrangians will lead to currents $K^a$ which satisfy $\partial_a (\sqrt{-g} K^a) = 0$ but the resulting $K^a$ will not be a generally covariant four-vector. Given the fact that, in GR, both
Einstein-Hilbert Lagrangian and the $\Gamma^2$ Lagrangian lead to such conserved currents shows that their difference --- which is purely a surface term --- will also lead to a conserved current. Therefore, in the context of GR, one can repeat the entire analysis using the current obtained from the surface term of Einstein-Hilbert Lagrangian.  Given the fact that the surface term in Einstein-Hilbert case is known to be closely related to horizon entropy, this fact de-mystifies the success of \va\ procedure to a limited extent.

\section{Cardy formula and entropy}

Using a suitably defined Fourier decomposition of the $T_1$ and $T_2$ in Eq.~(\ref{1.93}) and Eq.~(\ref{1.90}) we can find the central charge and zero mode eigenvalue, respectively. (The Fourier modes will have to be chosen such that the modes of the diffeomorphism generators satisfy Eq.~(\ref{Virasoro1}). The Cardy formula \cite{Cardy:1986ie,Carlip:1998qw} will then allow us to compute the entropy associated to the Killing horizon from the central charge and zero mode eigenvalue.
We start with a  Fourier decomposition of $T_1$ and $T_2$ given by:
\begin{eqnarray}
T_1 = \displaystyle\sum_m A_m T_m; \,\,\,\
T_2 = \displaystyle\sum_n B_nT_n~,
\label{Fourier}
\end{eqnarray}
with $A_n^* = A_{-n}$, $B_m^*=B_{-m}$ and the Fourier modes $T_m$ must be chosen in such a way that the Fourier modes of diffeomorphism generators $\xi_m$ satisfy Eq.~(\ref{Virasoro1}).  
Substituting Eq.~(\ref{Fourier}) in Eq.~(\ref{1.93}) we obtain:
\begin{eqnarray}
K[\xi_1,\xi_2]: = - \displaystyle\sum_{m,n} \frac{C_{m,n}}{32\pi G}\int \sqrt{h}d^{d-2}X P^{abcd}\mu_{ab}\mu_{cd}\frac{1}{\kappa} \Big(D T_m~D^2 T_n - D T_n~D^2 T_m\Big)~, 
\label{confsch2}
\end{eqnarray}
where $C_{m,n} \equiv A_mB_n$ and so $C_{m,n}^* = C_{-m,-n}$. Defining  the corresponding Fourier decomposition of 
\begin{equation}
K[\xi_1,\xi_2] = \sum_{m,n} C_{m,n} K[\xi_m,\xi_n]
\end{equation} 
 we find that:
\begin{eqnarray}
K[\xi_m,\xi_n]: =  - \frac{1}{32\pi G}\int \sqrt{h}d^{d-2}X P^{abcd}\mu_{ab}\mu_{cd} \frac{1}{\kappa} \Big(D T_m~ D^2 T_n - D T_n~D^2 T_m\big)~. 
\label{confsch}
\end{eqnarray}

To proceed further we need  to choose the explicit form of $T_m$. For a stationary space-times 
the coordinates near the horizon is chosen such that the (approximate) Killing vector $\chi^a$ is given by $\chi^a = (1,0,0,....)$, (It is possible to consider a more general case, suitable for stationary spacetimes with rotation; this is mentioned in Appendix \ref{AppendixC}). Then the usual ansatz for $T_m$ is:
\begin{eqnarray}
T_m = \frac{1}{\alpha}\exp\left[im\left(\alpha t  + g(x) + p.x_{\perp}\right)\right]~
\label{T}
\end{eqnarray}
where $\alpha$ is a constant and $g(x)$ is a  function that is regular  at the Killing horizon. $p$ is an integer and $x_{\perp}$ are the ($d-2$) tangential coordinates. Here $t-x$ plane defines the null surface.  This choice of overall normalisation automatically satisfies Eq.~(\ref{Virasoro1}) for any $\alpha$. This can be easily checked by expressing Eq.~(\ref{1.91}) in terms of the Fourier decomposition,
\begin{eqnarray}
\{\xi_1,\xi_2\}^a = \displaystyle{\sum_{m,n}}C_{m,n} \{\xi_m,\xi_n\}^a = \displaystyle{\sum_{m,n}}C_{m,n} \Big[\{T_m,T_n\}\chi^a + \{R_m,R_n\}\rho^a\Big]~.
\label{Fourier3}
\end{eqnarray}
Similar choice was made earlier in Ref. \cite{Silva:2002jq}. Interestingly, Eq.~(\ref{T}) is regular at the Killing horizon while the $T_m$ used  in \cite{Carlip:1999cy}, is not.
We can now compute the resulting \va, identify the central charge and compute the entropy. Obviously, the result will depend on the choice made for $\alpha$ and we need to fix this to get a unique value for entropy. A natural choice, arising from the fact that near-horizon Rindler geometry exhibits periodicity in imaginary time with period $2\pi/\kappa$, is
\begin{equation}
\alpha=\kappa
\label{condalpha}
\end{equation} 
However, we will postpone imposing this condition to the end and work with an arbitrary $\alpha$ in order to see the dependence of the Cardy entropy on $\alpha.$

Substituting Eq.~(\ref{T}) in Eq.~(\ref{confsch}) and defining a quantity
\begin{eqnarray}   
\hat{\cal{A}} =  - \frac{1}{2}\int \sqrt{h}d^{d-2}X P^{abcd}\mu_{ab}\mu_{cd}~,
\label{area}
\end{eqnarray}
which is
proportional to the Wald entropy,
we obtain,
\begin{eqnarray}
K[\xi_m,\xi_n]:
= -im^3\frac{\hat{\cal{A}}}{8\pi G} \frac{\alpha}{\kappa}\delta_{n+m,0}~.
\label{2.01}
\end{eqnarray}
(Note that $\hat{\mathcal{A}}$ reduces to the horizon area in the case of GR.).
 Similarly, using the Fourier decomposition, $Q[\xi] = \displaystyle\sum_{m} A_{m} Q[\xi_m]$ in Eq.~(\ref{1.90}), we obtain,
\begin{eqnarray}
Q[\xi_m] = \frac{\hat{\cal{A}}}{8\pi G}\frac{\kappa}{\alpha}\delta_{m,0}~.
\label{Q1}
\end{eqnarray}
Further, from Eq.~(\ref{1.92}), on using $Q[\{\xi_1,\xi_2\}] = \displaystyle\sum_{m,n} C_{m,n} Q[\{\xi_m,\xi_n\}]$, we can obtain the relation $Q[\{\xi_m,\xi_n\}] = -i(m-n)Q[\xi_{m+n}]$ where $Q[\xi_{m+n}]$ is given by Eq.~(\ref{Q1}).
Hence, Eq.~(\ref{1.77}) leads to:
\begin{eqnarray}
i[Q_m,Q_n] = (m-n)Q[\xi_{m+n}]  + m^3\frac{\hat{\cal{A}}}{8\pi G} \frac{\alpha}{\kappa}\delta_{n+m,0}~.
\label{Q2}
\end{eqnarray}
This is the 
standard form of the Virasoro algebra Eq.~(\ref{virasoro}) with $Q[\xi_{m+n}]\equiv Q_{m+n}$. We can identify the central charge and the zero mode eigenvalue  as:
\begin{equation}
\frac{C}{12} = \frac{\hat{\cal{A}}}{8\pi G}\frac{\alpha}{\kappa};
\qquad
Q[\xi_0] =  
 \frac{\hat{\cal{A}}}{8\pi G} \frac{\kappa}{\alpha}~.
\label{C}
\end{equation}
The standard Cardy formula for the entropy is given by \cite{Cardy:1986ie,Carlip:1998qw},
\begin{eqnarray}
S=2\pi\sqrt{\frac{C\Delta}{6}}~;\qquad \Delta \equiv Q_0-\frac{C}{24}
\label{cardy}
\end{eqnarray}
which leads to
\begin{eqnarray}
S=\frac{\hat{\cal{A}}}{4G}\Big[2-\frac{\alpha^2}{\kappa^2}\Big]^{\frac{1}{2}}
\to\frac{\hat{\cal{A}}}{4G}
\label{rindlerentropy}
\end{eqnarray}
This exactly matches with the Wald entropy if we take $\alpha=\kappa$. In the case of GR, we reproduce the Bekenstein-Hawking entropy. The motivation for the choice of  $\alpha=\kappa$ may be understood by introducing the Euclidean time which will briefly discussed in the next section.

\section{\label{Section6}Illustration: Rindler approximation in Riemann normal coordinates}

We will illustrate the analysis in the previous sections as well as some mathematical details of  Appendix \ref{AppendixB} in the simple context in this section. We consider an arbitrary event in a spacetime and introduce the Riemann normal coordinates around that event. We next boost to a local Rindler frame with acceleration parameter $\kappa$ in the x-direction which will introduce a local Rindler horizon as perceived by the accelerated observers. The  form of the metric near the horizon, in the ($x-t$) plane is given by,
\begin{eqnarray}
ds^2 = -\Big(2\kappa x +  Bx^2\Big)dt^2 + \frac{1}{2\kappa x} dx^2~.
\label{new1}
\end{eqnarray}
For this metric 
\begin{eqnarray}
\chi^a = (1,0);\,\,\,\ \chi_a= g_{ab}\chi^b = \Big(-(2\kappa x+ Bx^2),0\Big); \,\,\,\ \chi^2 = g_{ab}\chi^a\chi^b = -\Big(2\kappa x +  Bx^2\Big)
\label{new2}
\end{eqnarray}
showing that $x=\mathcal{O}(\chi^2)$ near the horizon.
Further from  \eq{1.121},
\begin{eqnarray}
\rho_a = \Big(0,\frac{1}{\kappa}(\kappa +  Bx)\Big);\,\,\,\ \rho^a = g^{ab}\rho_b = \Big(0,2(\kappa x +  Bx^2)\Big);\,\,\,\ \rho^2 = g_{ab}\rho^a\rho^b = \frac{2x}{\kappa}\Big(\kappa+ Bx\Big)^2~.
\label{new3}
\end{eqnarray}
The Killing horizon is given by ($\chi^2 = 0$)
\begin{eqnarray}
x = 0~,
\label{new4}
\end{eqnarray}
and the non-zero Christoffer connections are
\begin{eqnarray}
\Gamma^t_{tx} = \frac{\kappa+ Bx}{2\kappa x +  Bx^2};\,\,\,\ \Gamma^x_{tt} = 2\kappa x \Big( \kappa +  Bx\Big);\,\,\,\ \Gamma^x_{xx}=-\frac{1}{2x}~.
\label{new6}
\end{eqnarray}

  Now using the above values one can easily check that the left hand side of Eq.~(\ref{1.124}) is
\begin{eqnarray}
\rho^a\nabla_a T = 2\Big(\kappa x+ Bx^2\Big)\partial_x T~,
\label{new7}
\end{eqnarray}
which is by Eq.~(\ref{new2}) is of ${\cal{O}}(\chi^2)$. To illustrate the relation Eq.~(\ref{1.87}) we need to calculate left hand side and right hand side component by component. This leads to
\begin{eqnarray}
P^{abcd}\nabla_c\xi_d = P^{abcd}\xi_c\rho_d\frac{1}{\chi^2} (2\kappa T - \frac{1}{\kappa}D^2 T) + P^{abtx}\Big(-\frac{2\kappa +  Bx}{2\kappa(\kappa+ Bx)}\partial_t^2T + \frac{1}{\kappa^2}(\kappa+ Bx)\partial_t^2T\Big)
\label{new8}
\end{eqnarray}
in which the last term is of the ${\cal{O}}(\chi^2)$. Similarly Eq.~(\ref{1.80}) can be shown that it is exact upto ${\cal{O}}(\chi^2)$.

   Also, let us briefly discuss the assumptions and relations which are used in the main analysis to get the final expression. We first calculate $\frac{\rho^2}{\chi^2}$ for the above metric which upto order $x^2$ is given by
\begin{eqnarray}
\frac{\rho^2}{\chi^2} = -1 - \frac{2B}{\kappa^2}\Big(2\kappa x + \frac{1}{3}Bx^2\Big)~.
\label{new9}
\end{eqnarray}
  near the null surface, which has been used several times, automatically satisfied (see Eq.~(\ref{new2}) and Eq.~(\ref{new3})). Therefore $\frac{\rho^2}{\chi^2} = -1 + {\cal{O}}(\chi^2)$ and hence near the Killing horizon one can neglect the terms from ${\cal{O}}(\chi^2)$. A component wise verification will show that another assumption $\sigma^{ab}\nabla_b T =0$ for deriving the relation $2$ in Appendix \ref{AppendixB} is exactly satisfied for the above metric.

  Finally, consider the Euclideanised version ($t\rightarrow -i\tau$) of the metric near the horizon. In the Euclidean space our analysis still goes through with an  ansatz for $T_m$ taken as,
\begin{eqnarray}
T_m = \frac{1}{\alpha}e^{im(\alpha \tau + g(r) + p.x_{\perp})}~.
\label{new5}
\end{eqnarray}
In this case,  near the horizon $\frac{\rho^2}{\chi^2} =1 + {\cal{O}}(\chi^2)$. Following all the earlier steps one again obtains the same central term and the zero mode eigenvalue Eq.~ (\ref{C}). So the entropy will be Eq.~(\ref{rindlerentropy}). However, the Euclidean time $\tau$ \textit{must have}  the periodicity $\frac{2\pi}{\kappa}$, to avoid the conical singularity. Hence in  \eq{new5} we need to choose  $\alpha=\kappa$ to maintain this periodicity  in $\tau$.

\section{Conclusions}

The idea of obtaining horizon entropy from diffeomorphism generators near the horizon has a long history and has been attempted by several people using different techniques following the pioneering work by Carlip \cite{Carlip:1999cy}. All these approaches which we have referred to earlier do not always agree in the details or in the conceptual basis. All of them (as far as we know) were done on-shell with the equations of motion being used at one stage or the other. They also involve imposing certain boundary conditions or ignoring variations of certain terms in order to obtain the final result. Finally, all but the work in Ref.~\cite{Cvitan:2002rh} deals with Einstein's theory which, as we pointed out in section 1, is a bad discriminator of approaches to identify the entropy. In Einstein's theory horizon entropy \textit{happens} to be proportional to horizon area but not in more general \LL\ models.

In this paper we have attempted to overcome some of the limitations mentioned above. We have introduced a simple and physically meaningful definition for the variation of the current and the resultant bracket for the conserved charges. We did not require any specific boundary conditions in order to work out the central charge. More importantly, our entire analysis is off-shell and works for \LL\ models of gravity reproducing the correct Wald entropy for these models. We have also indicated how these ideas can work for any local Rindler horizon thereby connecting up with concepts in emergent gravity paradigm.  We believe this approach can possibly be further simplified and the mathematical details can be made more transparent.  We are now in the process of investigating these issues further.

Finally, it may be noted that our results add a different perspective to Virasoro algebra programme, which is possibly more in tune with gravity being an emergent phenomenon, in the following sense: In the more conventional approaches --- which uses charges defined on-shell and the field equations in the computations --- one thinks of the black hole horizon, say, as arising from a specific theory as a solution to the field equations and we obtain its entropy in the given theory. (We stress that the entropy of a black hole depends on the theory and is not simply proportional to horizon area in e.g., \LL\ models.). But in our approach, we only need the tensor $P^{abcd}$ (which has the symmetries of the curvature tensor and is divergence-free) to perform our computations and we get the entropy of the horizon to be the Wald entropy calculated using $P^{abcd}$. This is more in tune with the emergent, thermodynamic, perspective of gravity in which the entropy tensor $P^{abcd}$ is more fundamental. Just as thermodynamics of matter can be studied by extremising an entropy function $S(E,V)$, the dynamics of spacetime can be studied if the entropy tensor $P^{abcd}$ is given.
Mathematically, this arises because, once $P^{abcd}$ is given, one can associate a gravitational entropy $P^{ab}_{cd}\nabla_an^c \nabla_bn^d$ with the null vectors in spacetime;
it can be shown that \cite{aseemtp, paddyaspects} extremising the total entropy functional for all null vectors now leads to the field equations of \LL\ models. Therefore, in this perspective, we start with  $P^{abcd}$ which is fundamental; we then   determines the entropy density of spacetime and by  extremising it we obtain  the field equations. The entropy, in this sense, \textit{is} an off-shell construct and can be defined for any geometry if we are given a $P^{abcd}$. The fact that we can obtain the same entropy from Virasoro programme
working entirely off-shell, once $P^{abcd}$ is specified,  seems to be consistent with this picture.

\section*{Acknowledgement}

The research work of TP is partially supported by the J.C. Bose fellowship of Department of Science and Technology, India. 
The authors like to acknowledge Sanved Kolekar for useful discussions during the progress of the work.

\vskip 6mm
\section*{Appendices}
\appendix
\section{\label{AppendixA}($d-2$)-dimensional surface element}
\renewcommand{\theequation}{A.\arabic{equation}}
\setcounter{equation}{0}  
 In this Appendix we will give the expression for the surface element $\mu_{ab}$. The ($d-2$)-dimensional null surface we are interested in, is defined by $\chi^2=0$ 
where $\chi^a$ is the (approximate) Killing vector. If we introduce another auxillary null vector $N^a$  such that $\chi^aN_a = -1$ then  $d\Sigma_{ab} = d^{d-2}X\mu_{ab}$ where $\mu_{ab} = -(\chi_aN_b-\chi_bN_a)$. 
A convenient choice for $N^a$ can be obtained as follows: Let
$t^a$ be the tangent to the ($d-2$)-surface and $k^a$ be a null vector satisfying $k_a\chi^a = -1$, defined by 
\begin{eqnarray}
k^a = -\frac{1}{\chi^2}\Big(\chi^a - \frac{|\chi|}{\rho}\rho^a\Big)~,
\label{1.33}
\end{eqnarray}
Then we
define $N^a$ by  $N^a=k^a-\alpha\chi^a-t^a$. The condition $N^2 =0$ requires  $t^2 = -2\alpha-\alpha^2\chi^2$. So, to the leading order in $\chi^2$ we have
\begin{eqnarray}
\mu_{ab} = -\frac{|\chi|}{\rho\chi^2}(\chi_a\rho_b - \chi_b\rho_a)~.
\label{1.81}
\end{eqnarray}

\section{\label{AppendixB} Some important relations}
\renewcommand{\theequation}{B.\arabic{equation}}
\setcounter{equation}{0}  
    In this Appendix we will derive some important relations which are useful in the main calculation. The relations will be derived based on the relations given by Carlip (see Appendix A of \cite{Carlip:1999cy}). These are valid upto the leading order in $\chi^2$.

\subsection*{Relation 1:}   
   For a Killing vector $\chi^a$, which is null at the horizon, the rotation is given by,
\begin{eqnarray}
w_{ab} = \frac{1}{2}\Big(h^c_b\nabla_c\chi_a - h^c_a\nabla_c\chi_b\Big)
\label{1.31}
\end{eqnarray}
where the transverse metric $h^a_b$ is
\begin{eqnarray}
h^a_b = \delta^a_b + \chi^ak_b+k^a\chi_b~.
\label{1.32}
\end{eqnarray}
Now substituting (\ref{1.32}) in (\ref{1.31}) we obtain,
\begin{eqnarray}
w_{ab} = \frac{1}{2}\Big[-2\nabla_a\chi_b - \chi^ck_b\nabla_a\chi_c + k^c\chi_b\nabla_c\chi_a + \chi^ck_a\nabla_b\chi_c - k^c\chi_a\nabla_c\chi_b\Big]
\label{1.34}
\end{eqnarray}
Since,
\begin{eqnarray}
\chi^ck_b\nabla_a\chi_c = -\kappa k_b\rho_a~,
\label{1.35}
\end{eqnarray}
the above reduces to
\begin{eqnarray}
w_{ab} = \frac{1}{2}\Big[-2\nabla_a\chi_b + \kappa(k_b\rho_a - k_a\rho_b) + k^c\chi_b\nabla_c\chi_a - k^c\chi_a\nabla_c\chi_b\Big]~.
\label{1.36}
\end{eqnarray}
Using the expression for $k^a$ (\ref{1.33}) we obtain,
\begin{eqnarray}
k_b\rho_a = -\frac{1}{\chi^2} \Big(\chi_b\rho_a - \frac{|\chi|}{\rho}\rho_a\rho_b\Big)
\label{1.37}
\end{eqnarray}
and
\begin{eqnarray}
k^c\chi_b\nabla_c\chi_a = \frac{1}{\chi^2}\Big(-\kappa\chi_b\rho_a + \frac{|\chi|}{\rho}\rho^c\chi_b\nabla_c\chi_a\Big)~.
\label{1.38}
\end{eqnarray}
Substitution of these in (\ref{1.36}) yields,
\begin{eqnarray}
w_{ab} = \frac{1}{2}\Big[-2\nabla_a\chi_b + \frac{2\kappa}{\chi^2}(\chi_a\rho_b - \chi_b\rho_a) + \frac{|\chi|}{\chi^2\rho}(\rho^c\chi_b\nabla_c\chi_a - \rho^c\chi_a\nabla_c\chi_b)\Big]~.
\label{1.39}
\end{eqnarray}
Near the  horizon, where $w_{ab}\rightarrow 0$, we have,
\begin{eqnarray}
\nabla_a\chi_b =  \frac{\kappa}{\chi^2}(\chi_a\rho_b - \chi_b\rho_a) + \frac{|\chi|}{2\chi^2\rho}(\rho^c\chi_b\nabla_c\chi_a - \rho^c\chi_a\nabla_c\chi_b)~.
\label{1.40}
\end{eqnarray}
A solution for $\nabla_a\chi_b$ can be taken as
\begin{eqnarray}
\nabla_a\chi_b =  \frac{\kappa}{\chi^2}(\chi_a\rho_b - \chi_b\rho_a)~.
\label{1.41}
\end{eqnarray}
This can be verified by substituting this in Eq.~(\ref{1.40}). In this case
 $\rho^c\chi_b\nabla_c\chi_a - \rho^c\chi_a\nabla_c\chi_b = 0$. 
Furthermore, it can be verified component wise that the above is exact for the metric (\ref{new1}). Of course, if the metric coefficients contain the next leading order in $x$, then the relation Eq.~(\ref{1.41}) will have corrections to the order ${\cal{O}}(\chi^2)$ which in the near horizon limit can be neglected.  

\subsection*{Relation 2:}
Let us define a projection tensor
\begin{eqnarray}
\sigma^{ab} = g^{ab} - \frac{\chi^a\chi^b}{\chi^2} - \frac{\rho^a\rho^b}{\rho^2}~,
\label{1.42}
\end{eqnarray}
and assume that $T$ satisfies the condition $\sigma^{ab}\nabla_b T = 0$ near the horizon. This tells that the projection of $\nabla_aT$ along $\sigma$ is of ${\cal{O}}(\chi^2)$. Then
\begin{eqnarray}
\frac{\chi^a\chi^b}{\chi^2}\nabla_b T = \nabla^aT - \frac{\rho^a\rho^b}{\rho^2}\nabla_b T~.
\label{1.43}
\end{eqnarray}
The last term vanishes due to the boundary condition \eq{1.124}. Hence
\begin{eqnarray}
\nabla_aT = \frac{\chi_a}{\chi^2} D T
\label{1.44}
\end{eqnarray}
where $D=\chi^a\nabla_a$. We can ow obtain several further relations from these.
From (\ref{1.44}) we get,
\begin{eqnarray}
D(\nabla_a T) = D\Big[\frac{\chi_a}{\chi^2} D T\Big] = \frac{1}{\chi^2}\Big[\chi_aD^2T+\kappa\rho_aDT\Big]~.
\label{1.45}
\end{eqnarray}
Now use of (\ref{1.41}) and the condition (\ref{1.124}) yield,
\begin{eqnarray}
\nabla_a (DT) = \nabla_a(\chi^b\nabla_b T)
= - \frac{\kappa}{\chi^2}\rho_aDT + D(\nabla_aT)~.
\label{1.46}
\end{eqnarray}
Substituting (\ref{1.45}) in the above we obtain,
\begin{eqnarray}
\nabla_a (DT) = \frac{1}{\chi^2}\chi_aD^2T~.
\label{1.47}
\end{eqnarray}
Taking the covariant derivative of (\ref{1.44}) and then using (\ref{1.41}) and (\ref{1.47}) we have,
\begin{eqnarray}
\nabla_b\nabla_aT
&=& \frac{\kappa}{\chi^4}(\chi_b\rho_a - \chi_a\rho_b)DT + \frac{2\kappa}{\chi^4}\rho_b\chi_aDT + \frac{1}{\chi^4}\chi_a\chi_b D^2T
\nonumber
\\
&=& \frac{\kappa}{\chi^4}(\chi_b\rho_a + \chi_a\rho_b)DT + \frac{1}{\chi^4}\chi_a\chi_b D^2T
\label{1.48}
\end{eqnarray}

\subsection*{Relation 6:}
Consider the following linear combination form for $\nabla_a\rho_b$:
\begin{eqnarray}
\nabla_a\rho_b = A_1\chi_a\chi_b + A_2\rho_a\rho_b + A_3 \chi_a\rho_b + A_4\chi_b\rho_a + A_5g_{ab}+A_6\nabla_a\chi_b~.
\label{1.110}
\end{eqnarray}
This is justified because the calculation is near the null surface in the  ($t,x$) plane. 
Since, $\nabla_a\rho_b=\nabla_b\rho_a$, we have $A_3=A_4$ and $A_6=0$. Hence,
\begin{eqnarray}
\nabla_a\rho_b = A_1\chi_a\chi_b + A_2\rho_a\rho_b + A_3 (\chi_a\rho_b + \chi_b\rho_a) + A_5g_{ab}~.
\label{1.111}
\end{eqnarray}
Then use of the relation $\frac{\chi^a\chi^b}{\chi^2}\nabla_a\rho_b = -\frac{\kappa\rho^2}{\chi^2}$ (see Appendix A of \cite{Carlip:1999cy}) leads to,
\begin{eqnarray}
A_1\chi^2+A_5 = -\frac{\kappa\rho^2}{\chi^2}~.
\label{1.112}
\end{eqnarray}
Relation: $\rho^a\nabla_a\chi^b - \chi^a\nabla_a\rho^b=0$ [(A.2) of \cite{Carlip:1999cy}] yields,
\begin{eqnarray}
A_1\chi^2\chi_b + A_3\chi^2\rho_b + A_5\chi_b + \frac{\kappa\rho^2}{\chi^2}\chi_b = 0~,
\label{1.113}
\end{eqnarray}
where $\rho^a\nabla_a\chi^b$ is computed by using (\ref{1.41}).
Use of (\ref{1.112}) leads to $A_3 = 0=A_4$ and therefore,
\begin{eqnarray}
\nabla_a\rho_b = A_1\chi_a\chi_b + A_2\rho_a\rho_b + A_5g_{ab}~.
\label{1.114}
\end{eqnarray}
Now using $\frac{\rho^a\rho^b}{\rho^2}\nabla_a\rho_b = -\frac{\kappa\rho^2}{\chi^2} + {\cal{O}}(\chi^2)$ (equation A.8 of \cite{Carlip:1999cy}) we get,
\begin{eqnarray}
A_2\rho^2 + A_5 = -\frac{\kappa\rho^2}{\chi^2} + {\cal{O}}(\chi^2)~.
\label{1.115}
\end{eqnarray}
Again, $\nabla_a\rho^a = -\frac{2\kappa\rho^2}{\chi^2} + {\cal{O}}(\chi^2)$ (equation A.7 of \cite{Carlip:1999cy}) implies,
\begin{eqnarray}
A_1\chi^2 + A_2\rho^2 + d~A_5 = -\frac{2\kappa\rho^2}{\chi^2} + {\cal{O}}(\chi^2)~,
\label{1.116}
\end{eqnarray}
where $d$ is the spacetime dimension. Substituting (\ref{1.112}) and (\ref{1.115}) in the above and neglecting the $\chi^2$ order term, we obtain $A_5 = 0$. Therefore, (\ref{1.112}) and (\ref{1.115}) yield $A_1 = -\frac{\kappa\rho^2}{\chi^4}$ and $A_2 = - \frac{\kappa}{\chi^2}$, respectively. Hence (\ref{1.114}) reduces to 
\begin{eqnarray}
\nabla_a\rho_b = -\frac{\kappa}{\chi^2} \Big(\frac{\rho^2}{\chi^2}\chi_a\chi_b + \rho_a\rho_b \Big)~.
\label{1.71}
\end{eqnarray}
Since our analysis is near the event horizon where $\frac{\rho^2}{\chi^2} = - 1$, the above reduces to the following form,
\begin{eqnarray}
\nabla_a\rho_b = \frac{\kappa}{\chi^2} \Big(\chi_a\chi_b - \rho_a\rho_b \Big)~.
\label{1.72}
\end{eqnarray}
The above relation has corrections terms which are in ${\cal{O}}(\chi^2)$. This may be explicitly verified for the metric Eq.~(\ref{new1}).  
Finally using (\ref{1.41}) and (\ref{1.72}) we have,
\begin{eqnarray}
\nabla_d\nabla_a\rho_b = 0~.
\label{1.75}
\end{eqnarray}

\subsection*{Relation 7:}
Here, $R=\frac{\chi^2}{\kappa\rho^2}DT$. Therefore,
\begin{eqnarray}
\nabla_a R = \frac{1}{\kappa}\nabla_a(\frac{\chi^2}{\rho^2})DT + \frac{\chi^2}{\kappa\rho^2}\nabla_a(DT)~.
\label{1.117}
\end{eqnarray}
Use of (\ref{1.71}) gives $\nabla_a(\frac{\chi^2}{\rho^2}) = 0$ and hence (\ref{1.47}) leads to, 
\begin{eqnarray}
\nabla_a R = \frac{1}{\kappa\rho^2}\chi_aD^2T \simeq -\frac{1}{\kappa\chi^2}\chi_aD^2T~.
\label{1.118}
\end{eqnarray}

\subsection*{Relation 8:}
  From (\ref{1.118}) we obtain,
\begin{eqnarray}
\nabla_d\nabla_aR &=& -\frac{1}{\kappa}\Big[\nabla_d(\frac{1}{\chi^2})\chi_aD^2T + \frac{1}{\chi^2}(\nabla_d\chi_a)D^2T + \frac{1}{\chi^2}\chi_a\nabla_d(D^2T)\Big]
\nonumber
\\
&=& -\frac{1}{\kappa}\Big[\frac{\kappa}{\chi^4}(\chi_d\rho_a+\chi_a\rho_d)D^2T + \frac{1}{\chi^2}\chi_a\nabla_d(D^2T)\Big]~,
\label{1.50}
\end{eqnarray}
where (\ref{1.41}) have been used.
Now,
\begin{eqnarray}
\nabla_a(D^2 T) = 
 (\nabla_a\chi^b)\nabla_b(DT) + D[\nabla_a(DT)]~.
\label{1.102}
\end{eqnarray}
Using (\ref{1.41}) and (\ref{1.47}), the first term in the above reduces to,
\begin{eqnarray}
(\nabla_a\chi^b)\nabla_b(DT) = -\frac{\kappa}{\chi^2}\rho_aD^2T~,
\label{1.103}
\end{eqnarray}
while the last term gives,
\begin{eqnarray}
 D[\nabla_a(DT)] = \frac{\kappa}{\chi^2}\rho_aD^2T + \frac{1}{\chi^2}\chi_aD^3T~.
\label{1.104}
\end{eqnarray}
Substituting these in (\ref{1.102}),
we obtain,
\begin{eqnarray}
\nabla_a(D^2T) = \frac{1}{\chi^2}\chi_aD^3T~.
\label{1.51}
\end{eqnarray}
Therefore,
\begin{eqnarray}
\nabla_d\nabla_aR =  -\frac{1}{\chi^4}(\chi_d\rho_a+\chi_a\rho_d)D^2T - \frac{1}{\kappa\chi^4} \chi_a\chi_d D^3T~.
\label{1.52}
\end{eqnarray}

\subsection*{Relation 9:}
  Here, $\xi_a = T\chi_a + R\rho_a$. Hence
\begin{eqnarray}
\nabla_a\xi_b = \chi_b\nabla_a T + T\nabla_a\chi_b + \rho_b\nabla_aR +R\nabla_a\rho_b~.
\label{1.53}
\end{eqnarray}
Substitution of respective values in the above yields,
\begin{eqnarray}
\nabla_a\xi_b = \frac{\chi_a\chi_b}{\chi^2}DT + \frac{\kappa}{\chi^2}(\chi_a\rho_b - \chi_b\rho_a)T - \frac{1}{\kappa\chi^2}\chi_a\rho_bD^2T + R\nabla_a\rho_b~.
\label{1.54}
\end{eqnarray}
Hence,
\begin{eqnarray}
P^{abcd}\nabla_c\xi_d = P^{abcd}\Big[\frac{2\kappa}{\chi^2}T - \frac{1}{\kappa\chi^2}D^2T\Big]\chi_c\rho_d~.
\label{1.55}
\end{eqnarray}

\subsection*{Relation 10:}
From (\ref{1.41}),
\begin{eqnarray}
\nabla_c\nabla_a\chi_b = \frac{\kappa^2}{\chi^4}(\chi_a\rho_b - \chi_b\rho_a)\rho_c + \frac{\kappa}{\chi^2}(\chi_a\nabla_c\rho_b - \chi_b\nabla_c\rho_a)~.
\label{1.56}
\end{eqnarray}
Using (\ref{1.72}) we have 
\begin{eqnarray}
\chi_a\nabla_c\rho_b - \chi_b\nabla_c\rho_a = \frac{\kappa}{\chi^2} (-\chi_a\rho_b + \chi_b\rho_a)\rho_c~.
\label{1.73}
\end{eqnarray}
Hence, (\ref{1.56}) reduces to,
\begin{eqnarray}
\nabla_c\nabla_a\chi_b = 0~.
\label{1.74}
\end{eqnarray}

\subsection*{Relation 11:}
From (\ref{1.53}),
\begin{eqnarray}
\nabla_d\nabla_a\xi_b &=& (\nabla_d\chi_b)(\nabla_a T) + \chi_b\nabla_d\nabla_aT + (\nabla_dT)(\nabla_a\chi_b) + T\nabla_d\nabla_a\chi_b 
\nonumber
\\
&+& (\nabla_d\rho_b)(\nabla_a R)+\rho_b\nabla_d\nabla_aR+ (\nabla_dR)(\nabla_a\rho_b) + R\nabla_d\nabla_a\rho_b~.
\label{1.57}
\end{eqnarray}
Substituting the respective values we obtain,
\begin{eqnarray}
\nabla_d\nabla_a\xi_b = \frac{2\kappa}{\chi^4}\chi_a\rho_b\chi_dDT - \frac{1}{\kappa\chi^4}\chi_a\rho_b\chi_dD^3T-\frac{1}{\chi^4}\chi_a\chi_b\chi_dD^2T~.
\label{1.76}
\end{eqnarray}
This final expression was given in \cite{Cvitan:2002rh} without the details of the derivation. Here we gave the details for the shake of completeness and we belief that it will help to the reader for the future purpose.

\section{\label{AppendixC}  Inclusion of rotational terms}
\renewcommand{\theequation}{C.\arabic{equation}}
\setcounter{equation}{0}  
For a stationary space-times with rotation, 
the coordinates near the horizon can be chosen such that the (approximate) Killing vector $\chi^a$ is given by $\chi^a = (1,0,0,\Omega_1,\Omega_2,....)$, $\Omega_j$ are the rotational parameters. Let $\Omega\equiv\sum\Omega_j$. Then the  ansatz for $T_m$ generalises to:
\begin{eqnarray}
T_m = \frac{1}{N}\exp\left[im\left(\alpha t + \sum\Phi_j + g(x,\theta)\right)\right]~
\equiv\frac{1}{N}\exp\left[im\left(\alpha t + \Phi + g(x,\theta)\right)\right]~,
\label{T1}
\end{eqnarray}
where $\alpha$ is a constant, $\Phi_j$'s are the coordinates on which the metric does not depend on, $\Phi\equiv\sum\Phi_j$  and $g(x,\theta)$ is a  function that is regular  at the Killing horizon.   This choice satisfies Eq.~(\ref{Virasoro1}) with $N = (\alpha + \Omega)$. 
Similar choice was made earlier in Ref. \cite{Silva:2002jq}. 
The limits of the integration are chosen such that $T_n$ has periodicity $2\pi$ and $\pi$ on $\Phi_j$ and $\theta$, respectively. 
The rest of the analysis proceeds exactly as in the main text.

\end{document}